\documentclass[twocolumn,aps,floats,longbibliography,superscriptaddress]{revtex4-1}
\usepackage[latin9]{inputenc}
\usepackage{amsmath}
\usepackage{amssymb}
\usepackage{graphicx}
\usepackage{wasysym}
\PassOptionsToPackage{normalem}{ulem}
\usepackage{ulem}
\usepackage[unicode=true,pdfusetitle,
 bookmarks=true,bookmarksnumbered=false,bookmarksopen=false,
 breaklinks=false,pdfborder={0 0 1},backref=false,colorlinks=false]
 {hyperref}
\hypersetup{colorlinks,linkcolor=blue,citecolor=blue,urlcolor=blue}

\makeatletter

\usepackage{xcolor}
\usepackage{pdfcolmk}
\usepackage{wasysym}

\usepackage{txfonts}

\makeatother
\begin{document}
\title{Featureless quantum paramagnet with frustrated criticality and competing spiral magnetism
\\ on spin-1 honeycomb lattice magnet}
\author{Jian Qiao Liu}
\affiliation{International Center for Quantum Materials, Peking University, Beijing, China}
\affiliation{Department of Physics and HKU-UCAS Joint Institute 
for Theoretical and Computational Physics at Hong Kong, 
The University of Hong Kong, Hong Kong, China}
\author{Fei-Ye Li}
\affiliation{Department of Physics and HKU-UCAS Joint Institute 
for Theoretical and Computational Physics at Hong Kong, 
The University of Hong Kong, Hong Kong, China}
\affiliation{State Key Laboratory of Surface Physics and Department of Physics,
Fudan University, Shanghai, 200433, China}
\author{Gang Chen}
\email{gangchen.physics@gmail.com}
\affiliation{Department of Physics and HKU-UCAS Joint Institute 
for Theoretical and Computational Physics at Hong Kong, 
The University of Hong Kong, Hong Kong, China}
\affiliation{State Key Laboratory of Surface Physics and Department of Physics,
Fudan University, Shanghai, 200433, China}
\affiliation{Collaborative Innovation Center of Advanced Microstructures, 
Nanjing University, Nanjing, 210093, China}
\author{Ziqiang Wang}
\affiliation{Department of Physics, Boston College, Chestnut Hill, Massachusetts 02467, USA}

\date{\today}
\begin{abstract}
We study the spin-1 honeycomb lattice magnets with frustrated exchange interactions. 
The proposed microscopic spin model contains first and second neighbor Heisenberg 
interactions as well as the single-ion anisotropy. 
We establish a rich phase diagram that includes a featureless quantum paramagnet
and various spin spiral states induced by the mechanism of order by quantum disorder. 
Although the quantum paramagnet is dubbed featureless, it is shown that, 
the magnetic excitations develop a contour degeneracy in the reciprocal space
at the band minima. These contour degenerate excitations are responsible for the 
frustrated criticality from the quantum paramagnet to the ordered phases. 
This work illustrates the effects of magnetic frustration on both magnetic orderings 
and the magnetic excitations. We discuss the experimental relevance to various 
Ni-based honeycomb lattice magnets.    
\end{abstract}

\maketitle

\section{Introduction}



Frustrated magnetism is a large topic in modern strongly correlation physics~\cite{PhysRev.79.357,Vannimenus_1977,Balents}. 
Frustration usually refers to competing interactions that cannot be optimized simultaneously. 
For magnetic systems, these interactions are the exchange interactions between the local magnetic 
moments. The Ising interaction on the triangular lattice is often used as an example to illustrate
the concept of frustration~\cite{PhysRev.79.357}. Due to the strong magnetic frustration, 
conventional magnetic orders are often expected to be suppressed. Instead, unconventional 
quantum phases such as quantum spin liquids~\cite{Savary2016,Lee_2007,RevModPhys.89.025003}, 
skyrmion lattices and spin nematics~\cite{Fert2017,Kohama10686,Zhitomirsky2010}, 
and exotic excitations such as spinons, topological magnons~\cite{FeiYe2016,Kondo_2019,2018NatPh1011Y,2018NatCoSong} 
and magnetic monopoles~\cite{Castelnovo,PhysRevB.69.064404,PhysRevB.96.195127}, may emerge.  
As an important notion in modern condensed matter physics, magnetic frustration does not seem to have 
a strong phenomenological correspondence or a mathematical characterization. 
This is quite different from other contemporary notions such as emergent symmetry~\cite{2004Sci1490S}, 
topology~\cite{RevModPhys.82.3045,RevModPhys.83.1057,2012arXiv1210.1281W} 
and entanglements~\cite{Laflorencie}. By comparison, magnetic frustration 
is more like the physical origin or the driving force for unconventional 
magnetic properties, rather than a description of the internal structures and the physical consequences. 
Thus, we are more interested in the understanding of various physical consequences that result
from the magnetic frustration in magnetic systems. 

Empirically, magnetic frustration could lead 
to a large number of degenerate or nearly degenerate low-energy states such that
the system has a difficulty to develop a conventional magnetic order and the ordering
temperature is often suppressed. An empirical parameter, dubbed ``frustration parameter''~\cite{Balents},
is thus used to characterize the level of frustration of the system. 
The ``frustration parameter'' is defined as the ratio between the Curie-Weiss temperature
and the ordering temperature. The larger the ``frustration parameter'' is, the more frustrated
the system is. This simple empirical parameter, however, does not actually provide much 
information or understanding about the low-energy physical properties of the system. 
For the low-energy physical properties, one can then focus on the ground state phases
and the corresponding emergent physics associated with the quantum phases.
These emergent physics are usually controlled by the quantum phases rather 
than being connected to the magnetic frustration in a direct fashion. 
Therefore, it seems that the magnetic frustration merely works as an 
empirical route to look for unconventional quantum phases with spin degrees of freedom.  
In this paper, we work on the specific spin-1 honeycomb magnet
and show that, the emergent low-energy magnetic properties are 
directly tied to the magnetic frustration in this system. 

\begin{figure}[b]
	\includegraphics[width=8.5cm]{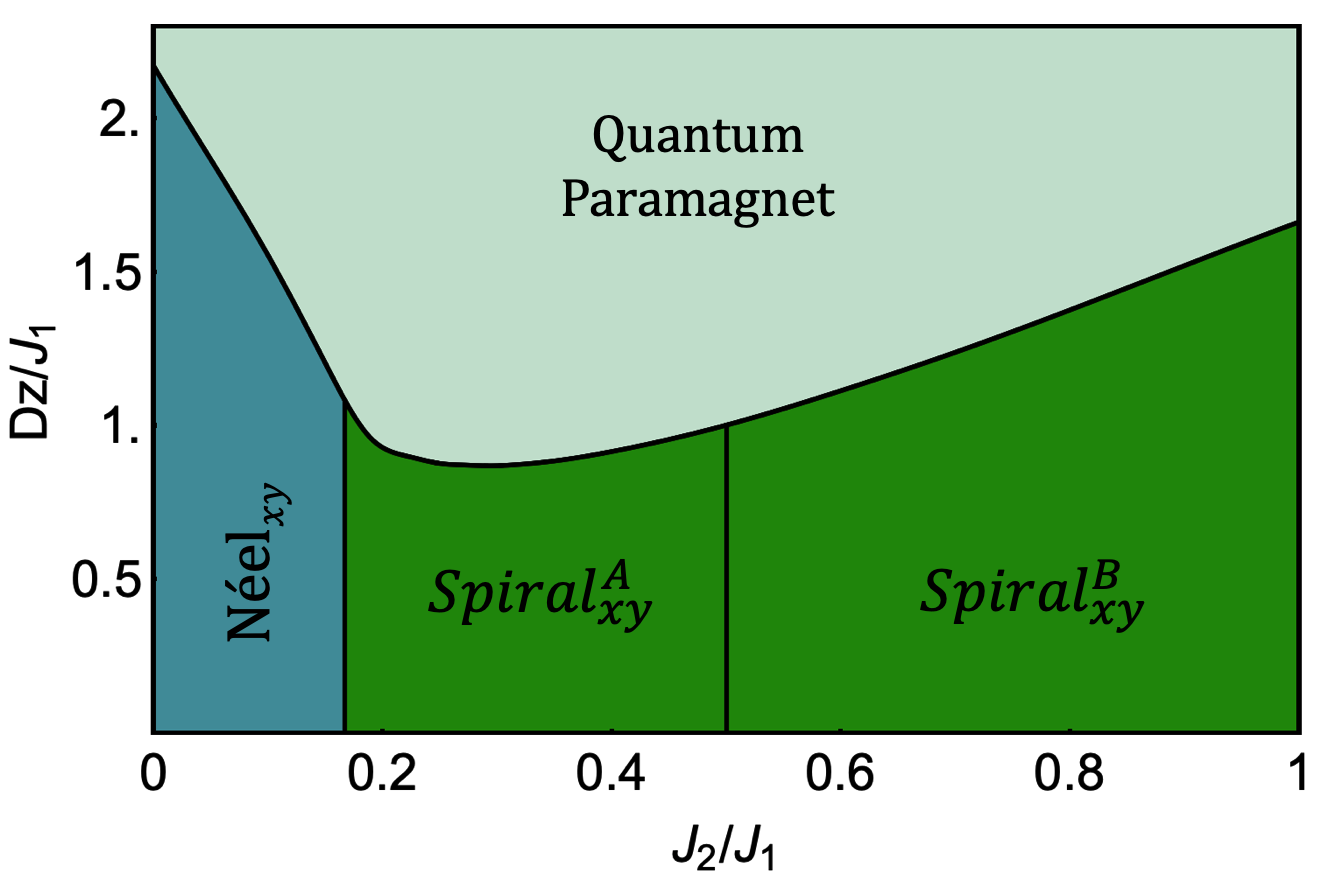}
	\caption{Ground state phase diagram of the $J_1$-$J_2$-$D_z$ 
model. The details of each phase are explained in the main text.}
	\label{fig1}
\end{figure}

Spin-1/2 honeycomb lattice magnets, especially the honeycomb 
Kitaev materials~\cite{2017arXiv170107056T,Hermanns,2019NatRPTakagi}, 
have attracted a lot of attention in the field. 
The spin-orbital entanglement of the local moments brings 
rather anistropic interactions between neighboring spins 
and creates a strong frustration and a disordered Kitaev 
spin liquid ground state even for a geometrically 
unfrustrated honeycomb lattice~\cite{2006AnPhy}. Recent efforts try to 
extend spin-1/2 Kitaev materials to high-spin magnets, mainly spin-1 magnets,
where the heavy ligand atoms may bring some anisotropic spin
interactions through the exchange path~\cite{PhysRevLett.123.037203}. 
Although exact solutions are not available for high spin Kitaev 
materials, it is hoped that,
exotic quantum state may still persist to high spin systems, especially
since quantum effect in spin-1 magnets is still strong and 
magnetic frustration could further enhance it. 
Moreover, it is well-known that, with larger physical Hilbert space, 
the spin-1 magnet could bring more distinct physics from the spin-1/2 
magnet. This has been well illustrated in the strong (topological) 
distinction between the spin-1/2 Heisenberg chain and spin-1 Haldane
chain~\cite{Haldane}. 
Compared to the tremendous efforts in various spin-1/2 magnets, the 
attention in spin-1 magnets is rather limited. Thus,
it is particularly timing to explore the potentially rich physics of 
frustrated spin-1 magnets~\cite{PhysRevLett.109.016402,PhysRevB.96.020412,PhysRevB.98.045109,PhysRevLett.108.087204,2015arXiv151101505S}. 

On the experimental side, several new materials have been proposed as 
spin-1 honeycomb magnets and show distinct magnetic properties. 
All of them are Ni-based and have either a honeycomb layer structure
or a buckled honeycomb structure. Inspired by the growing interest in 
the spin-1 magnets and the existing experiments on spin-1 honeycomb 
magnets, we propose and study a minimal model for spin-1 local moments
on a honeycomb lattice with the Hamiltonian,
\begin{eqnarray}
H= J_1 \sum_{\langle ij \rangle} {\boldsymbol S}_i \cdot {\boldsymbol S}_j 
+  J_2 \sum_{\langle\langle ij \rangle \rangle} {\boldsymbol S}_i \cdot {\boldsymbol S}_j 
+ D_z \sum_{i} (S^z_i)^2,
\label{ham}
\end{eqnarray}
where ${\boldsymbol S}_i$, ${\boldsymbol S}_j$ are spin-1 local moments 
on the honeycomb lattice, $\langle ij \rangle$ denotes exchange interactions 
of the first neighbor spin pairs, $\langle \langle ij \rangle\rangle$ 
denotes exchange interactions of the second neighbor
pairs. 
We work on the regime with ${J_1 >0}$ and ${J_2 >0}$, and a 
ferromagnetic $J_1$ is obtained by applying time reversal 
transformation on one sublattice. 
Here we are mostly interested in the Heisenberg spin degrees of freedom. 
We introduce a single-ion spin anisotropy that is generically allowed by the 
planar lattice geometry and the large spin moment. This single-ion spin anisotropy
is necessary for the materials that show a clear in-plane and out-plane spin anisotropy. 
We are interested in the easy-plane regime with ${D_z > 0}$ for the model. 
It is ready to see that the easy-axis regime is connected to the Ising limit 
where the quantum effect can be suppressed. In our analysis, we start from the 
well-defined limit with a strong easy-plane anisotropy. In this limit, the 
ground state is clearly known as a trivial and featureless quantum paramagnet
with ${|S^z_i=0\rangle}$ on each site. 
As we show in this paper, although this quantum paramagnet is dubbed
``trivial and featureless'', the magnetic frustration brings extra features
on top of this featureless ground state. The coherent magnetic excitations 
develop a contour degenerate band minima when the frustration of the exchange 
interaction becomes large. We identify this property as one direct
consequence of the magnetic frustration. 
As the single-ion anisotropy becomes weaker, the gap of the magnetic excitation 
is reduced and eventually becomes zero, and the system develops magnetism. 
It is the degenerate low-enegy magnetic excitations that are responsible for the 
critical fluctuations and the development of the magnetism. We further show the 
unusual critical properties due to these critical modes in the vincinity of the 
phase transition between the featureless quantum paramagnet and the ordered phases.
This is again linked to the magnetic frustration.  
On the ordered side, the ordering structure of the system is directly related to the 
degenerate low-energy band minima in the quantum paramagnet and the critical modes. 
We show, quantum fluctuations are needed to break the degeneracy among the candidate 
ordering wavevectors on the degenerate contours in the momentum space. The 
consequence of this order by quantum disorder is explained. 
The full phase diagram of the model is summarized in Fig.~\ref{fig1}.

The remaining part of the paper is organized as follows. 
In Sec.~\ref{sec2}, we study the featureless quantum paramagnet 
with a large single-ion anisotropy and point out the nontrivial 
features in the magnetic excitations. 
In Sec.~\ref{sec3}, we consider the instabilities of the featureless 
quantum paramagnet and study the critical properties of the frustrated 
quantum criticality. 
In Sec.~\ref{sec4}, we focus on the ordered regime and explain the 
magnetic orders from the mechanism of order by quantum disorder. 
Finally in Sec.~\ref{sec5}, we discuss various experimental relevance 
to the Ni-based honeycomb lattice magnets. 

\section{``Featureless'' quantum paramagnet}
\label{sec2}


\begin{figure*}[t]
 \includegraphics[width=18cm]{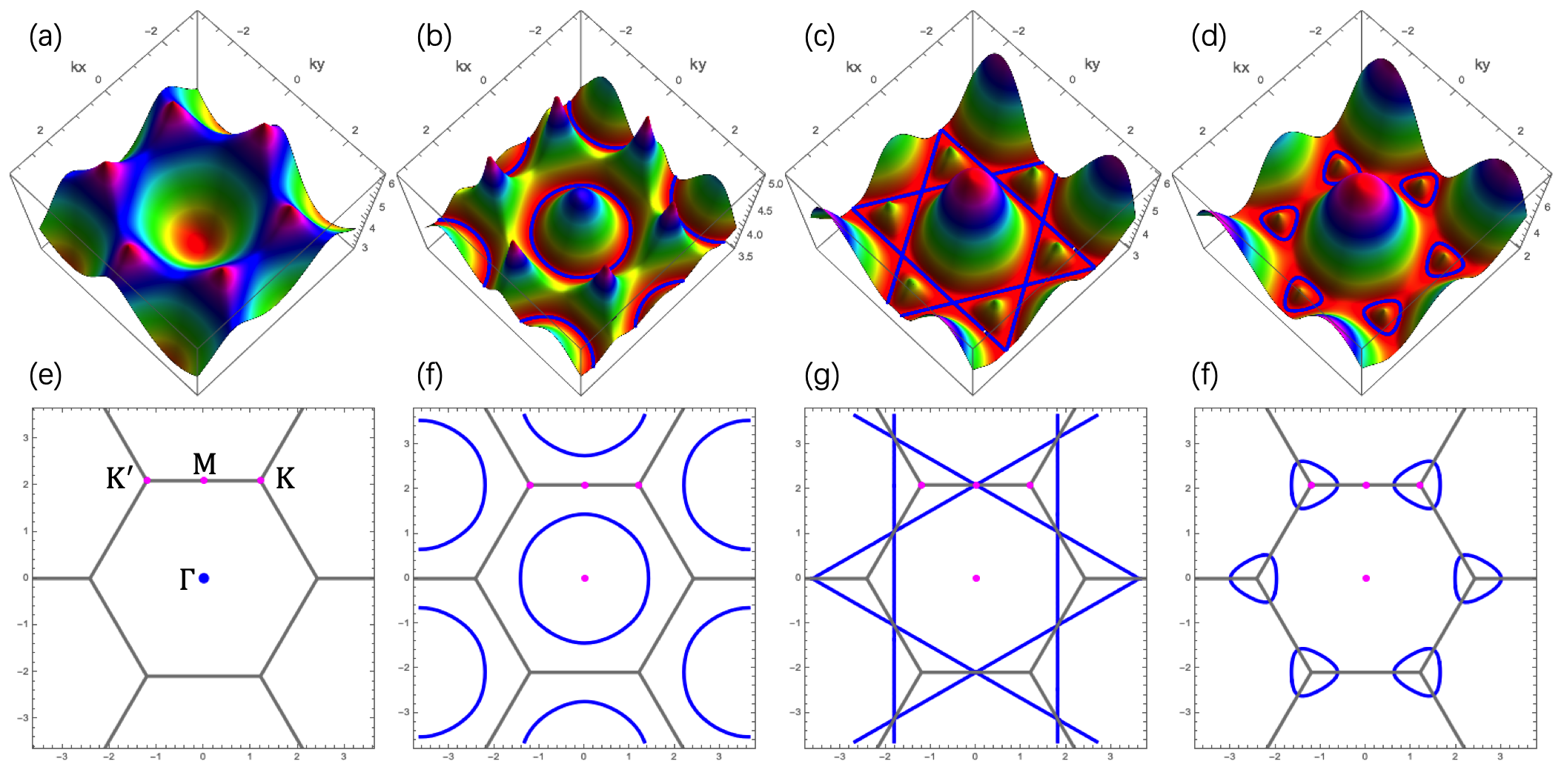}
 \caption{{Upper panel}: The magentic excitation $\omega_- (\mathit{k})$  
  in the $k_x-k_y$ plane of the quantum paramagnet with different parameters.
  {Lower panel}: The degenerate excitation minima (blue) depicted in the Brillouin Zone
  (gray). High-symmetry points are marked as pink.  We set ${D_z=6J_1}$ in the figure, 
  and $J_2/J_1$ takes the following parameters (a,e) ${J_2=0.1J_1}$; (b,f) ${J_2=0.3J_1}$;
  (c,g) ${J_2 =0.5J_1}$; (d,h) ${J_2 =0.7J_1}$.}
  \label{fig2}
\end{figure*}

We start from the strong single-ion limit where there is a well-defined ground 
state to work with. When $D_z \gg J_1, J_2$, the ground state is 
a trivial quantum paramagnet and is approximated by
\begin{eqnarray}
\left|\Psi_\text{GS}\right\rangle = \prod_{i} \left|S_i^z \equiv 0\right\rangle. 
\label{eq2}
\end{eqnarray}
It is well-known that this trivial state has no order of any kind and 
preserves all the symmetries of the original Hamiltonian. Thus, it is
simply featureless. As we will show below, however, the magnetic
excitation of this featureless state develops some extra features 
when the system becomes frustrated. This property may be interpreted
as the physical consequence of the frustration on this featureless state. 
Since this state does not have any conventional magnetic order,
the conventional spin wave theory cannot be applied to compute 
the magnetic excitation with respect to this featureless state. 
Here, we adopt the flavor wave theory~\cite{PhysRevLett.81.3527,PhysRevB.60.6584,PhysRevB.98.045109}.
This theory was originally developed for the interacting spin-orbital local moments 
with an effective fundamental SU(4) representation on each site of the triangular lattice.
This theory is not only applicable to the featureless ground state
for the spin-1 magnets, but also applies to any kind of product states 
where the ground state can be separated into the direct product of 
local states for local sites or local cluster units. 

The spirit of the flavor wave theory is similar to the Schwinger boson parton
representation of the spin variables. Here, one introduces three
boson operators for the three states of the spin-1 Hilbert space
and condenses one of them to generate the featureless quantum paramagnetic
state. Henceforth, the remaining two auxiliary bosons describe the magnetic 
excitations. 
The advantange of the flavor wave approach is that the physical nature of the 
ground state and the corresponding excitations is straight-forward. 
For this purpose, we define the mapping between 
the boson operators and the spin states as 
 \begin{align}
a_{i,1}^\dagger |\emptyset\rangle  &\equiv |S_i^z=1\rangle, \nonumber \\
a_{i,0}^\dagger |\emptyset\rangle  &\equiv |S_i^z=0\rangle, \nonumber \\
a_{i,\bar{1}}^\dagger |\emptyset\rangle  &\equiv |S_i^z=\bar{1}\rangle,
 \end{align}
where the state $|\emptyset\rangle$ is the vacuum state.
The physical spin-one operator, $S^\alpha_i$, can be written as
\begin{eqnarray}
S^\alpha_i \equiv 
\sum_{m,n}\big\langle S_i^z=m\big| S^\alpha_i \big|S_i^z=n \big\rangle a_{i,m}^\dagger a^{}_{i,n},
\end{eqnarray}
 with $\alpha=x,y,z$ and $m,n=1,0,\bar{1}$.
 Apparently, this mapping enlarges the physical Hilbert space.
Thus we need to impose a constraint to get back to the physical Hilbert space with 
    \begin{align}
 a_{i,1}^\dagger a_{i,1}^{\phantom\dagger} + 
 a_{i,0}^\dagger a_{i,0}^{\phantom\dagger} + 
 a_{i,\bar{1}}^\dagger a_{i,\bar{1}}^{\phantom\dagger} 
 \equiv 1. 
 \end{align}
on each lattice site. In terms of these flavor wave bosons, the 
exchange part of Eq.~\eqref{ham} is converted into four-boson interacting
terms while the single-ion anisotropy is quadratic in the boson
operators. 
To obtain the quantum paramagnetic state in Eq.~\eqref{eq2}, one
simply condenses the $a_0$ boson by replacing
 \begin{align}
 \langle a_{i,0}^\dagger \rangle \approx \langle a_{i,0}^{\phantom\dagger} \rangle \rightarrow
 ({1-a_{i,1}^\dagger a_{i,1}^{\phantom\dagger} - a_{i,\bar{1}}^\dagger a_{i,\bar{1}}^{\phantom\dagger}})^{\frac{1}{2}}
 \end{align}
in the Hamiltonian. At the level of linear flavor wave treatment, 
one keeps the quadratic part of the bosonic operators in the 
Hamiltonian. This linear flavor wave Hamiltonian is given as 
 \begin{align}
 H_\text{fw}=\sum_{\bm{k} \in \text{BZ}} 
 \begin{pmatrix}
 \psi_{\bm{k},\text{A}}^\dagger & \psi_{\bm{k},\text{B}}^\dagger
 \end{pmatrix}
 \mathcal{M}(\bm{k})
 \begin{pmatrix}
 \psi_{\bm{k},\text{A}}^{\phantom\dagger} \\ \psi_{\bm{k},\text{B}}^{\phantom\dagger}
 \end{pmatrix},
 \label{fwham}
 \end{align}
 where ${ \psi_{\bm{k},\mu} \equiv \Big(a_{\bm{k},1,\mu}^{\phantom\dagger},
 a_{\bm{k},\bar{1},\mu}^{\phantom\dagger},
 a_{\bm{\bar{k}},1,\mu}^\dagger,
 a_{\bm{\bar{k}},\bar{1},\mu}^\dagger\Big)^\text{T} }$
 and ${\mu=A,B}$ labels the two sublattices of the honeycomb lattice. 
The Hamiltonian matrix $\mathcal{M}(\bm{k})$ is given as
 \begin{align}
\mathcal{M}(\bm{k})=\begin{pmatrix}
\mathcal{M}_1 & \mathcal{M}_2 \\
\mathcal{M}_2^* & \mathcal{M}_1
\end{pmatrix},
\end{align}
where
\begin{align}
\mathcal{M}_1&=
\begin{pmatrix}
m_2 & 0 & 0 & m_2 \\
0 & m_2 & m_2 & 0 \\
0 & m_2 & m_2 & 0 \\
m_2 & 0 & 0 & m_2
\end{pmatrix}
+D_z \mathcal{I}_{4\times4}, \\
\mathcal{M}_2&=\begin{pmatrix}
m_1 & 0 & 0 & m_1 \\
0 & m_1 & m_1 & 0 \\
0 & m_1 & m_1 & 0 \\
m_1 & 0 & 0 & m_1
\end{pmatrix},
\end{align}
and $\mathcal{I}_{4\times4}$ is a $4\times 4$ identity matrix. Here the 
entries of the matrix are 
\begin{align}
m_1\equiv& J_1 \sum_{\mu}
e^{-i{\bm{k}}\cdot\bm{b}_{\mu}}, \\	
m_2\equiv& J_2 \sum_{\mu}
e^{-i{\bm{k}}\cdot\bm{d}_{\mu}},
\end{align} 
where the summations above are over the first-neighbor 
vectors $\{ \bm{b}_{\mu} \}$ and the second-neighbor vectors $\{\bm{d}_{\mu} \}$
of the honeycomb lattice, respectively.

Using the Bogoliubov transformation, we establish the magnetic excitations
from the linear flavor wave Hamiltonian. The dispersions are 
 \begin{align}
 \omega_{\pm}(\bm{k})=\sqrt{D_z \left( D_z+2\left[J_2(\Lambda^2(\bm{k})-3) 
         \pm J_1\Lambda(\bm{k})\right]\right)},
 \label{fw_dispersion}
 \end{align}
where $\Lambda(\bm{k})\equiv\big|\sum_{\mu}
e^{-i{\bm{k}}\cdot\bm{b}_{\mu} }\big|$, with $\{\bm{b}_{\mu}\}$ 
being three nearest neighbor vectors of the honeycomb lattice.
Here both two branches $\omega_{+}(\bm{k})$ and $\omega_{-}(\bm{k})$ 
are two-fold degenerate, that is associated to the time reversal symmetry,
i.e., the equivalence between the ${S_z=1}$ excitation and the ${S_z= -{1}}$ 
excitation. In total we have ${4=2\times2}$ branches of magnetic excitations, 
that is consistent with the number of the sublattices and the flavors.
This is quite different from the coherent spin wave excitations
for a conventional magnetically ordered state where the branch number 
is equal to the number of the magnetic sublattices. 

On the top panel of Fig.~\ref{fig2}, we depict the evolution of the band structure 
for the low-lying mode $\omega_{-}(\bm{k})$ in the reciprocal space by varying 
$J_2/J_1$. As expected, the flavor wave magnetic excitations are fully gapped. 
A further examination of the spectrum points to a contour degeneracy at the 
band minimum. To understand that, we realize that the band minimum 
of $\omega_{-}(\bm{k})$ occurs at
  \begin{align}
\Lambda(\bm{k})=\frac{J_1}{2J_2},
 \label{spiral}
 \end{align}
and the solutions of the above equation determine the 
positions of the band minima. When ${J_2/J_1\leq 1/6}$, 
a single band minimum is realized at the $\Gamma$ point. 
When ${J_2/J_1>1/6}$, degenerate minima with a 
contour degeneracy in the reciprocal space are realized.
As $J_2/J_1$ increases from 1/6, 
 the contour first emerges surrounding the $\Gamma$ point, and gradually expands.
 At ${J_2/J_1=1/2}$, the contour becomes a perfect hexagon that 
 is formed by connecting the six equivalent M points at the Brillouin zone boundary. 
 As $J_2/J_1$ is further increased, the contour surrounds $K$ and $K^\prime$ points, 
 and finally the contour shrinks to $K$ and $K^\prime$ points when ${J_2/J_1 \rightarrow \infty}$. 
 The evolution of the degenerate contour is depicted on the lower panel of Fig.~\ref{fig2}.
The emergence of this degenerate contour arises from the frustration that 
is introduced by the competing $J_1$-$J_2$ interaction. 
With only first neighbor $J_1$ interaction, the exchange part is simply the 
nearest-neighbor Heisenberg model on the bipartite honeycomb lattice and is thus not frustrated. 
With only second neighbor $J_2$ interaction, the exchange part is the 
nearest-neighbor Heisenberg model on the triangular lattice of each 
sublattice and is known to be not very frustrated. In the intermediate $J_2/J_1$ and when 
$J_2$ and $J_1$ are comparable, a large frustration is expected for the exchange part,
and thus we experience a contour degenerate band minima in the excitation spectra.
This is a consequence of magnetic frustration on the quantum mechanical excitations 
of a featureless quantum paramagnetic state.

\section{Frustrated quantum criticality}
\label{sec3}

Inside the quantum paramagnetic phase, the interesting aspect occurs 
in the magnetic excitation spectrum. As the single-ion anisotropy decreases,
the exchange part of the interaction becomes more important and controls
the ground state properties of the system. 
The transition out of the quantum paramagnetic state can be traced 
by examining the excitation spectrum. As the parameter $D_{z}$ 
becomes smaller, the gap of the excitation diminishes.
At the critical value of $D_z$, the gap of the excitation becomes zero 
and the corresponding lowest energy mode starts to be condensed. 
This picture works very well for $J_2/J_1\leq 1/6$ where the 
flavor bosons are condensed at the $\Gamma$ point and the system
develops an antiferromagnetic order that preserves the translational
symmetry. 
The transition is a conventional 2+1D XY transition. 
For the regime with ${J_2/J_1 > 1/6}$, the flavor bosons have 
a difficulty to find an ordering wavevector to be condensed 
because of the contour degeneracy for the lowest energy modes. 
Although the high-order interactions between the flavor bosons
would eventually break the degeneracy and select the condensed 
mode, the presence of the contour degeneracy at the lowest energy 
modes should control the low-energy physics in the vincinity of the phase
transition from the quantum paramagnet. At the phase transition,
all these degenerate modes at the bottom of the band become
critical at the same time. Thus, this is a different kind of 
critical physics from the conventional one where only 
discrete numbers of bosonic modes become critical. 
From the excitation spectra, the low-lying modes
have no dispersion on the contour but disperse
linearly in the momentum direction normal to the 
contour. This critical property of the bosons 
near the degenerate contour looks a bit similar to the 
fermion modes near the Fermi surface in two spatial
dimensions, and we have a constant density of states at 
low energies. Thus, when thermal fluctuations are included 
in the critical regime, we would expect a 
linear-$T$ heat capacity just like what a Fermi surface 
would do. In the following, we demonstrate this result explicitly. 

The flavor wave theory applies well to the zero-temperature excitations
deep inside the quantum paramagnet,
but has an obstacle to extend to finite temperatures. Moreover, the 
renormalization or correction from the high-order interactions may become
important when $D_z$ gets closer to the critical points and the quantum
paramagnet phase becomes unstable. To capture the thermal fluctuation in the 
critical regime, we turn to a different approach. Because our Hamiltonian
has a global U(1) symmetry, we can map the spin variables into equivalent rotor 
variables. We introduce an integer-valued operator $n_i$ and the 
$2\pi$-periodic phase variable $\varphi_i$ such that $[\varphi_i,n_j] = i\delta_{ij}$.
We further identify $S_i^z$ as $n_i$, and $S^\pm_i$ as $\sqrt{2} e^{\pm i \varphi_i}$. 
Under this identification, we have actually allowed $n_i$ to take all integer values 
instead of $0,\pm 1$ for $S^z_i$. This extension would not cause any significant
effects as the weights of higher integer values are strongly suppressed by the 
single-ion anisotropy in the Hamiltonian. The spin Hamiltonian now becomes 
\begin{eqnarray}
H_{\text{rotor}} &=& \sum_{\langle ij \rangle} 
J_1 \big[ 2\cos( \varphi_i - \varphi_j)  + n_i n_j \big]  \nonumber \\
&+& \sum_{\langle\langle ij \rangle\rangle} 
J_2 \big[ 2\cos( \varphi_i - \varphi_j)  + n_i n_j \big]  \nonumber \\
&+& \sum_i D_z n_i^2. 
\end{eqnarray}
To compute the excitation spectra and obtain the dynamical properties, we implement 
the coherent state path integral formulation and formally integrate out the 
variable $n_i$. The resulting partition function is 
\begin{eqnarray}
\mathcal{Z} = \int \mathcal{D} \Phi \, \mathcal{D} \lambda
\, \exp \big[{-\mathcal{S} - i \sum_i \int_\tau (|\Phi_i|^2 -1 )} \big],
\end{eqnarray}
where the action $\mathcal{S}$ is given as 
\begin{eqnarray}
\mathcal{S} &\equiv& \int_\tau 
\sum_{\bm{k}}  \partial_{\tau}^{} \Phi^{\ast}_{\mu,\bm{k}} 
(4D_z \mathcal{I}_{2\times 2} + 2 \mathcal{J}_{\bm{k}})^{-1}_{\mu\nu} 
\partial_{\tau}^{} \Phi^{ }_{\nu,\bm{k}} 
\nonumber \\
&+& \sum_{i,j} J_{ij} \Phi_i^{\ast} \Phi_j .
\end{eqnarray}
Here $J_{ij}$ is the exchange matrix in the position space, and 
$J_{ij} \equiv J_1$ ($J_2$) if $ij$ is the first (second) neighbor. 
$\mathcal{J}_{\bm{k}}$ is the exchange matrix in the reciprocal 
space. As there are two sublattices, $\mathcal{J}_{\bm{k}}$ is 
a $2\times 2$ matrix and $\mu,\nu$ are sublattice indices. 
$\mathcal{I}_{2\times2}$ is a $2\times 2$ identity matrix. 
The complex field $\Phi_i$ is identified as $e^{i\varphi_i}$,
and we have the constraint $|\Phi_i| =1$ on each site that 
is enforced by the Lagrange multiplier $\lambda_i$.  
To solve for the dynamics, we implement an usual saddle
point approximation for the path integral. As the translation
symmetry is preserved both in the quantum paramagnetic phase
and at the finite temperatures, it is legitimate to assume
an uniform ansatz for the Lagrange multiplier $\lambda_i$
with $i\lambda_i\equiv \beta {\Delta}(T)$ at the saddle point.
By integrating out the $\Phi$ field, we obtain the saddle point equation,
\begin{eqnarray}
\frac{1}{N}\sum_{i=\pm} \sum_{\bm{k}} \frac{2D_z + \xi_i (\bm{k})}{\omega_{i}(\bm{k},T)} 
\coth [ \frac{\beta \omega_{i}(\bm{k},T) }{2} ] = 1,
\label{saddlepoint}
\end{eqnarray}
where $N$ is the total number of the lattice sites, $\omega_i(\bm{k},T)$
is the excitation spectrum and has a temperature dependence, 
and $\xi_i(\bm{k})$ is the eigenvalue of the exchange matrix $\mathcal{J}_{\bm{k}}$. 
Here we have 
\begin{eqnarray}
\xi_i(\bm{k}) &=& J_2 \sum_{\mu} \cos (\bm{k}\cdot \bm{d}_{\mu}) 
 \pm \big|J_1  \sum_{\mu}  \exp({i \bm{k} \cdot \bm{b}_{\mu}}) \big| 
 \nonumber \\
 &=& J_2 \big[  \Lambda^2(\bm{k}) -3 \big] \pm J_1 \Lambda (\bm{k})
\end{eqnarray}
with $\{ d_{\mu}\}$ the six vectors connecting second-neighbor sites,
and 
\begin{equation}
\omega_i(\bm{k},T) =\sqrt{ \big[4D_z + 2 \xi_i (\bm{k}) \big]\big[\Delta(T) + \xi_i (\bm{k}) \big]}. 
\end{equation}
The parameter $\Delta(T)$ is solved from the saddle point equation 
in Eq.~\eqref{saddlepoint} for each temperature. Just like the 
zero-temperature excitation of the quantum paramagnet in the previous section,
the low-lying excitation $\omega_-(\bm{k},T)$ develops the same contour degeneracy 
at the band minima for the same choice of $J_2/J_1$. 

In the paramagnetic phase both for the 
finite temperature and zero temperature, the $\Phi$ field is 
not condensed, and we do not need to single out the condensed 
piece in Eq.~\eqref{saddlepoint}. At the zero-temperature
phase transition, the spectrum becomes gapless with
\begin{equation}
\Delta ({T=0}) \equiv 3J_2+ \frac{J_1^2}{4J_2}. 
\end{equation}
The critical $D_z$ at the transition is obtained  
by solving the saddle point equation, and we establish
the phase boundary between the quantum paramagnet 
and the ordered ones in the zero-temperature phase
diagram of Fig.~\ref{fig1}. 
The critical $D_z$ at the phase boundary is non-monotonic
as one increases $J_2/J_1$ from 0, and becomes minimal 
at intermediate $J_2/J_1$. This indicates
the maximal frustration at the intermediate $J_2/J_1$ regime.  

To reveal the critical property in the regime with $J_2>J_1/6$ where
a contour degenerate critical modes exist,  
we tune the single-ion anisotropy
to the criticality and analyze the saddle point equation. 
At ${D_z = D_{zc}}$ and finite temperatures, $\Delta(T)$ has 
a temperature dependence and is defined as
\begin{eqnarray}
\Delta (T) &\equiv & \Delta (T=0) + \Delta'(T) \equiv
\Delta_0 + \Delta'(T).
\end{eqnarray}
The gapless low-lying excitation 
picks up a self-energy from $\Delta'(T)$ and has the form
near the band bottom 
\begin{eqnarray}
\omega_- (\bm{k},T) &\approx & \sqrt{ (4 D_{zc} -2 \Delta_0) \Delta'(T) 
+ v^2_{\perp,\bm{k}_c} \bm{k}_{\perp}^2 } \nonumber \\
& \equiv & 
\sqrt{ 2A \Delta'(T) 
+ v^2_{\perp,\bm{k}_c} \bm{k}_{\perp}^2 } ,
\end{eqnarray}
where we have made a Taylor's expansion at the band bottom, 
$\bm{k}_c$ is the momentum running along the degenerate contour,  
 $\bm{k}_{\perp}$ is the momentum component normal to the tangent 
direction of the degenerate contour
and $v_{\perp,\bm{k}_c}$ is the corresponding speed.
The weak temperature dependence of $v_{\perp,\bm{k}_c}$ has been
neglected in this expansion. The saddle point equation can be 
decomposed as 
\begin{eqnarray}
c 
+ \int_{ \bm{k}_c, \bm{k}_{\perp} }^{\Gamma}  
\frac{ A \coth \Big[ \frac{\beta}{2} \sqrt{ 2A\Delta'(T)+ v^2_{\perp,\bm{k}_c} \bm{k}_{\perp}^2}\Big]  }
{ \sqrt{ 2A\Delta'(T)+ v^2_{\perp,\bm{k}_c} \bm{k}_{\perp}^2}} =1,
\end{eqnarray}
where the momentum integration is over the area surrounding
the degenerate contour, the constant $c$ is 
an approximately temperature independent contribution from 
the integration outside this area and from the $\omega_+$ branch. 
At low temperatures, the temperature dependent part of the 
integration becomes independent of the momentum cutoff $\Gamma$, 
and depends on $T$ through the dimensionless parameter $A\Delta'(T)/T^2$.
In order for the integration to be constant in the temperature such that  
the saddle point equation is satisfied, we expect $\Delta'(T)\rightarrow T^2$ 
and hence $C_v \sim T$ in the limit $T\rightarrow 0$ under this analysis.

Here we make a further remark on the critical property. 
The enhanced density of the low-energy states is induced 
by the contour degeneracy for ${J_2>J_1/6}$.  
Since the contour degeneracy is accidental due to the 
frustrated interaction, there is no symmetry protection 
for the contour degeneracy. The contour degeneracy will 
eventually be lifted by fluctuations beyond the analysis above, 
leading to a modified specific heat behavior at the zero-temperature 
limit, but this may be difficult to be observed in experiments. 
For ${J_2 \leq J_1/6}$, a single critical mode implies a conventional 
$C_v \propto T^2$ behavior up to a correction from the fluctuations 
beyond the current analysis. If one deviates from the 
critical point, several crossovers and/or transition will 
appear as one increases the temperatures. This behavior 
will be discussed in Sec.~\ref{sec5}.



\section{Spiral magnetism from quantum order by disorder}
\label{sec4}

\begin{figure}[t]
	\includegraphics[width=5.5cm]{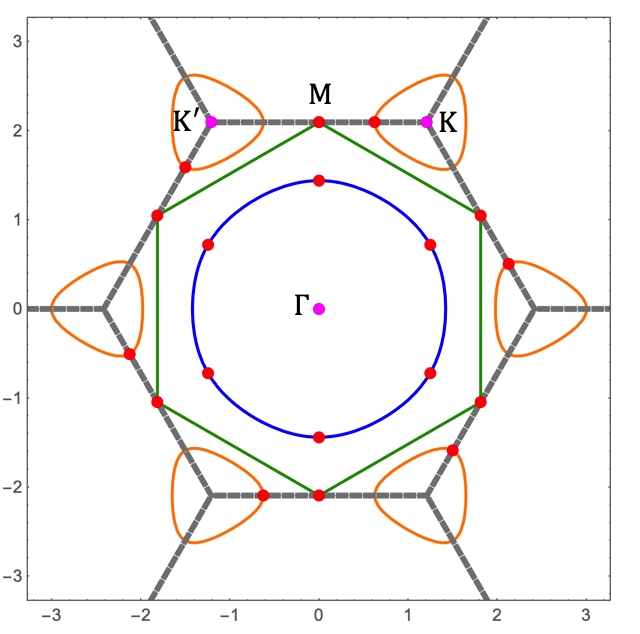}
	\caption{The evolution of the degenerate contour for the 
	candidate spiral wavevectors
	 for ${J_2/J_1=0.3,0.5,0.7}$ from inside to outside.
	The red points indicate the ordering wavevectors selected by quantum order by disorder. 
	The Brillouin zone boundary is shown in gray. }
	\label{contour}
\end{figure}

When the exchange part of the Hamiltonian becomes important, the system 
will eventually develop magnetism. In this regime, we will show that 
the magnetic frustration plays the traditional role just like what 
it is conventionally thought. As we expect magnetism, it is legitimate
to introduce well-defined order parameters to study the ground state 
magnetic properties with a traditional Weiss type of mean-field theory. 
One could further find the magnetic ordering structures by treating the 
spin operators as classical vector order parameters and optimizing the 
mean-field energy. The single-ion anisotropy with a positive $D_z$ favors 
the spins to be ordered in the $xy$ plane. We thus model the magnetic 
order parameter as 
\begin{eqnarray}
\langle {\boldsymbol S}_i \rangle &\equiv & m \big[  \cos (\bm{q}\cdot \bm{r}_i) \hat{x} 
                                                   + \sin (\bm{q}\cdot \bm{r}_i) \hat{y}\, \big] ,
                                   \label{spirala}\\
\langle {\boldsymbol S}_j \rangle &\equiv & m \big[  \cos (\bm{q}\cdot \bm{r}_i + \theta_{\bm{q}}) \hat{x} 
                                           + \sin (\bm{q}\cdot \bm{r}_i + \theta_{\bm{q}} ) \hat{y}\, \big] \label{spiralb},       
\end{eqnarray}
for $i \in$ A sublattice and $j \in$ B sublattice, respectively. Here the 
offset phase $\theta_{\bm{q}}$ between two sublattices depends explicitly 
on the spiral wavevector $\bm{q}$. The order parameter $m$ depends on the 
ratio between the single-ion anisotropy and the exchange interaction. For 
the continuous transition at $D_{zc}$, the order parameter $m$ increases
gradually from 0 and becomes maximal at ${D_z=0}$. In contrast, the ordering
wavevector $\bm{q}$ is decided by the ``kinetic part'', {\sl i.e.} the exchange
interactions. Thus, the order parameter $m$ and the ordering wavevector $\bm{q}$
are separately optimized or determined. It is ready to obtain that
the exchange interaction, at the mean-field level, requires the ordering
wavevector to satisfy the following conditions.  
For ${J_2<J_1/6}$, a simple antiferromagnetic N\'{e}el state is expected 
with ${\bm{q}=0}$ and ${\theta_{\bm{q}} = \pi}$. This state is labelled
as N\'{e}el$_{\text{xy}}$ state in the phase diagram of Fig.~\ref{fig1}.  
For ${J_2>J_1/6}$, there exists a degenerate contour in the reciprocal space
where the optimal $\bm{q}$'s reside. Remarkably, this degenerate contour for 
the ordering wavevector $\bm{q}$ is precisely the degenerate contour that is 
formed by the band minima of the excitation spectrum in the quantum paramagnetic 
phases, and we have
\begin{eqnarray}
\Lambda ({\bm{q}}) &\equiv& \Big|\sum_{\mu} \exp [-i \bm{q}\cdot \bm{b}_{\mu}] \Big| 
  =\frac{J_1}{2J_2} , \label{spiralq} \\
  \theta_{\bm{q}} & = & \pi + \arg\Big(\sum_{\mu} \exp [-i \bm{q}\cdot \bm{b}_{\mu}]\Big). 
\end{eqnarray}
Here the contour degeneracy is not protected by any symmetry of the system 
and thus is an artifact of the mean-field treatment. It is expected that,
once quantum fluctuations beyond the mean-field theory are included, the 
contour degeneracy will be lifted. This effect is known as quantum order by 
disorder. To establish the breaking of contour degeneracy, we implement a 
standard linear spin wave analysis for the candidate spiral magnetic orders
with a finite ordering wavevector $\bm{q}$. This approach not only produces 
the magnetic ordering structures but also generates the spectrum of the magnetic 
excitations.

\begin{figure*}[htb]
	\includegraphics[width=18cm]{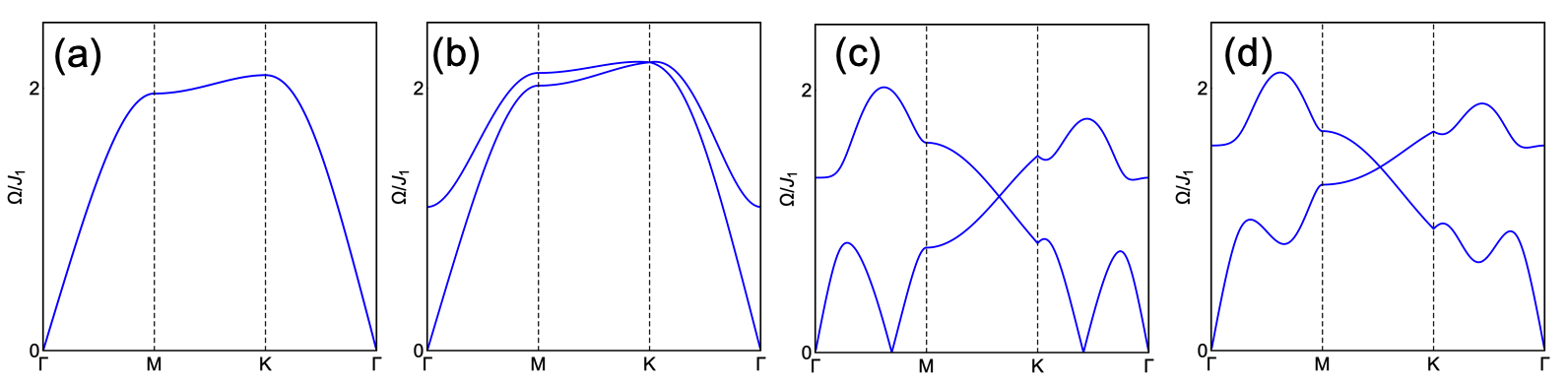}
	\caption{Representative spin wave excitations along high symmetry momentum lines. 
	(a) ${J_2=0.1J_1}$, ${D_z=0}$; (b) ${J_2=0.1J_1}$, ${D_z=0.1J_1}$;
	(c) ${J_2=0.3J_1}$, ${D_z=0}$; (d) ${J_2=0.3J_1}$, ${D_z=0.1J_1}$. 
	For (c) and (d), we choose the spin configuration
	with the wavevector $\boldsymbol{Q}_A$ from Eq.~\eqref{Qa}.}
	\label{fig4}
\end{figure*}

For the spin spiral order that is defined in Eqs.~\eqref{spirala} and ~\eqref{spiralb},
we rewrite the spin operators by introducing the following Holstein-Primakoff
bosons,
\begin{eqnarray}
&&{\boldsymbol S}_i\cdot\hat{n}_i =1-b_i^\dagger b_i^{\phantom\dagger}, \\
&&{\boldsymbol S}_i\cdot\hat{z}   =\frac{\sqrt{2}(b_i^{\phantom\dagger}+b_i^\dagger)}{2}, \\
&&{\boldsymbol S}_i\cdot(\hat{z}\times\hat{n}_i) =\frac{\sqrt{2}(b_i^{\phantom\dagger}-b_i^\dagger)}{2i},
\end{eqnarray}
where $\hat{n}_i$ is the orientation of the spin order at site $i$. 
With this substitution of the spin operators, we obtain the leading
spin wave correction to the classical ground state energy,
\begin{widetext} 
\begin{align}
H_{\text{sw}}=\frac{1}{2}\sum_{\bm{k}\in \text{BZ}}
(b_{\bm{k}\text{A}}^\dagger,b^{}_{\bm{\bar{k}}\text{A}},
b_{\bm{k}\text{B}}^\dagger,b^{}_{\bm{\bar{k}}\text{B}})
\begin{pmatrix}
M_1 & M_2 \\
M_2^* & M_1 
\end{pmatrix}
\begin{pmatrix} 
b_{\bm{k}\text{A}}^{} \\ b_{\bm{\bar{k}}\text{A}}^\dagger \\ 
b_{\bm{k}\text{B}}^{} \\ b_{\bm{\bar{k}}\text{B}}^\dagger
\end{pmatrix}
+C+E_{cl},
\end{align}
where the matrixes and constants are given as
\begin{align}
&M_1= \frac{J_2}{2}\sum_{\mu} e^{i\bm{k}\cdot\bm{d}_{\mu}}
\begin{pmatrix} 
1+\cos\Theta_{\bm{d}_{\mu}} & 1-\cos\Theta_{\bm{d}_{\mu}} \\ 
1-cos\Theta_{\bm{d}_{\mu}}  &1+ \cos\Theta_{\bm{d}_{\mu}}  \\
\end{pmatrix} 
  - \mathcal{I}_{2\times 2} \Big(J_1 \sum_{\mu} \cos\Theta_{\bm{b}_\mu} 
+ J_2 \sum_{\mu} \cos\Theta_{\bm{d}_\mu}\Big) 
+ D_z \begin{pmatrix} 1 & 1 \\1 & 1 \end{pmatrix}, \\
&M_2=\frac{J_1}{2} \sum_{\mu} e^{i\bm{k}\cdot\bm{b}_\mu}
\begin{pmatrix} 
1+ \cos\Theta_{\bm{b}_\mu} &1- \cos\Theta_{\bm{b}_\mu} \\ 
1-\cos\Theta_{\bm{b}_\mu}& 1+\cos\Theta_{\bm{b}_\mu} \\
\end{pmatrix}, \\
&C=\frac{N}{2}
\Big(J_1 \sum_{\mu} \cos\Theta_{\bm{b}_\mu}+ J_2 \sum_{\mu} 
\cos\Theta_{\bm{d}_\mu}\Big),\\
&E_{cl} = \frac{N}{2}\Big( J_1 \sum_\mu \cos \Theta_{\bm{b}_\mu}
+ J_2 \sum_\mu \cos \Theta_{\bm{d}_{\mu}} \Big),
\end{align}
with $\Theta_{\bm{b}_\mu} \equiv \bm{q}\cdot\bm{b}_\mu+\theta_{\bm{q}}$,
$\Theta_{\bm{d}_\mu }\equiv \bm{q}\cdot\bm{d}_\mu$,
and $\sum_\mu$ are over the first neighbor vectors $\{b_\mu\}$ or the second 
neighbor vectors $\{d_\mu\}$ of the honeycomb lattice.
Here $E_{cl}$ is the classical ground state energy of the spiral ordered state.
Diagonalizing this linear spin-wave Hamiltonian via a generalized Bogoliubov
transformation, we obtain two spin-wave modes $\Omega_{\pm}(\bm{k})$.
Accordingly, the zero-point energy is given as
\begin{align}
E_\text{zpt}=\frac{1}{2}\sum_{\bm{k}\in \text{BZ}} \sum_{\pm} \Omega_{\pm}(\bm{k})
+C+E_{cl} . 
\end{align}
As expected, the quantum energy correction lifts the contour degeneracy 
when ${J_2>J_1/6}$, and exhibits discrete band minima at certain modes.
The optimal spiral wavevectors are given by
\begin{eqnarray}
&&{\boldsymbol Q}_A =\left(0, \frac{2}{3} \cos^{-1}\left[\left(\frac{J_1}{4J_2}\right)^2\ 
-\frac{5}{4}\right]\right), \quad\quad\quad\quad\quad \text{for} \, {\frac{J_1}{6}<J_2<\frac{J_2}{2} },
\label{Qa}\\
&&{\boldsymbol Q}_B=\left(\frac{2}{\sqrt{3}}\cos^{-1} 
\left(\frac{J_1}{4J_2}+\frac{1}{2}\right), \frac{2\pi}{3} \right),
\quad\quad\quad\quad\quad  \text{for} \,  {{J_2}>\frac{J_1}{2}}, 
\label{Qb}
\end{eqnarray}
\end{widetext} 
and their symmetry equivalents.
As long as ${D <D_{zc}}$, the exact form of classical degenerate contour
as well as the quantum selected wavevectors, only depend on the value of
$J_2/J_1$. As we plot in Fig.~\ref{contour}, we illustrate the evolution 
of degenerate contour and optimal wavevectors as increasing $J_2/J_1$ 
from 1/6. When ${J_2/J_1 >1/6}$, the contour emerges, surrounds the $\Gamma$
point, and gradually expands; it then touches the boundary of the
Brillouin zone when ${J_2/J_1=1/2}$; as $J_2/J_1$ is further increased,
the contour surrounds $K$ and $K^\prime$ points, and finally it
shrinks to $K$ and $K^\prime$ points when ${J_2/J_1\to \infty}$,
corresponding to the $120^\circ$ order on two decoupled triangular
lattices. Our results in this section are consistent with the previous studies on the 
honeycomb lattice $J_1$-$J_2$ model without the anisotropy for the generic spins~\cite{PhysRevB.81.214419}
where the contour degeneracy and order-by-disorder were established at ${D_z=0}$.
The positive anisotropic $D_z$ term in our model simply makes the $xy$ plane as 
an easy plane to set spins on. The discrete wavevectors selected by the quantum 
zero point energy are marked by red points in Fig.~\ref{contour}. The resulting 
two different spiral states characterized by $\boldsymbol{Q}_A$ and 
$\boldsymbol{Q}_B$ are denoted as $\text{Spiral}_{xy}^A$ and
$\text{Spiral}_{xy}^B$ in Fig.~\ref{fig1}, respectively.

Once the magnetic order is determined from the quantum order by disorder, we 
proceed to evaluate the magnetic excitation with respect to the ordered states. 
In Fig.~\ref{fig4}, we show the magnetic excitations in spin ordered states.
For the N\'eel$_{xy}$ phase, the spin wave excitations are two-fold degenerate 
when ${D_z=0}$, with two Goldstone modes at the $\Gamma$ point, see Fig.~\ref{fig4}(a).
The presence of $D_z$ splits the degeneracy, and only one Goldstone mode is left
at the $\Gamma$ point, reflecting the symmetry reducing from SU(2) to U(1), 
see Fig.~\ref{fig4}(b). Similar effects occur for the Spiral$_{xy}$ phase, 
except the relevant Goldstone modes appear at both the $\Gamma$ point and 
the ordering wavevector, see Fig.~\ref{fig4}(c,d).

\section{Discussion} 
\label{sec5}

Here we discuss various aspects that are related to actual experiments. 
Probably the most clear indication for the presence of the single-ion 
anisotropy appears in the magnetic susceptibilities of the single-crystal 
samples. If one applies the field normal to the honeycomb plane and 
in the honeycomb plane, our 
simple mean-field calculation gives two different Curie-Weiss
temperatures with
\begin{eqnarray}
\Theta^z_{CW} &=& -D_z/3 - (2J_1+4J_2) ,\\
\Theta^\perp_{CW} &=& + D_z /6 - (2J_1+4J_2).
\end{eqnarray}
The comparison between these two temperatures could actually 
approximately give the value of the single-ion anisotropy. 

As our model exhibits a global U(1) symmetry, in principle there
would exist a Berezinsky-Kosterliz-Thouless transition out of the 
ordered phases as one increases temperature. However, the actual situation 
in realistic materials is more complex. The interlayer coupling and 
other anisotropic interactions would intervene and disrupt or alter 
this transition. Due to the presence of the contour degeneracy on the ordered 
sides 
for ${J_2 >J_1/6}$, when the thermal fluctuation smears the difference
in the quantum zero point energy for the spiral states from the 
degenerate contour, a crossover to the quantum critical regime of the
frustrated quantum criticality occurs and we expect the results obtained
in Sec.~\ref{sec3}.
When thermal fluctuations weaken the quantum effect, the thermal 
fluctuation would be dominated by the fluctuation modes near the degenerate
contour at low temperatures on the ordered side, this paramagnetic regime
is sometimes referred as ``spiral spin liquid''. It is not a new phase but a 
thermal regime with interesting equal-time spin correlation properties due to the degenerate manifold
that governs the low-temperature fluctuations. Here it is a 2D version,
in contrast to the 3D version with a surface degeneracy on the diamond lattice in Ref.~\onlinecite{Bergman}. 
 On the quantum paramagnetic side, the crossover 
temperature to the critical regime is set by the gap of the magnetic excitations. 

A series of materials have been proposed as spin-1 honeycomb lattice magnets~\cite{PhysRevB.96.214428,REGNAULT19801021,REGNAULT1986329,PhysRevB.93.024412,PhysRevMaterials.3.014410,2019JMMM481100W,Seibel}. 
These include the layer honeycomb material BaNi$_2$(XO$_4$)$_2$ (X = V, P, As), 
the buckled honeycomb material Ba$_2$NiTeO$_6$,  
Na$_3$Ni$_2$BiO$_6$, Li$_3$Ni$_2$SbO$_6$, Na$_3$Ni$_2$SbO$_6$
and Ni$_2$Mo$_3$O$_8$. All of them are Ni-based. 
The compounds BaNi$_2$(XO$_4$)$_2$ (X = V, P, As) have been found to 
have a strong frustration as well as an easy-plane anisotropy. 
In particular, BaNi$_2$V$_2$O$_8$ is a quasi-2D material and 
develop a N\'eel order. The magnetic excitation has very similar 
magnetic spectra as shown in Fig.~\ref{fig4}(b). While the difference 
exists in the lowest mode, the spectrum of BaNi$_2$V$_2$O$_8$ 
detected by inelastic neutron scattering has two gapped modes, 
rather than one gapped and one gapless mode shown in Fig.~\ref{fig4}(b).
The gap in the lower mode is explained by the additional easy-axis anisotropy 
within the honeycomb plane. The magnitude of this easy-axis anisotropy is
four order times smaller than the magnitude of exchange interactions such
that the spectrum from the inelastic neutron scattering is 
consistent with our prediction except the small gap in the lowest mode.
The easy-axis anisotropy breaks the continuous U(1) symmetry down to 
$\mathbb{Z}_2$ such that the system can develop a long-range magnetic 
order with a finite temperature phase transition. 

In the buckled honeycomb magnet Ba$_2$NiTeO$_6$~\cite{PhysRevB.93.024412}, the Ni$^{2+}$ ions are    
arranged in the A-B stacking patten and each layer is a triangular 
lattice. The A-B stacked bilayer is equivalent to a honeycomb lattice. 
The collinear order that was discovered by neutron diffraction in 
Ref.~\onlinecite{PhysRevB.93.024412} is equivalent to the N\'eel state in 
Fig.~\ref{fig1}. More experiments can be performed on this compound 
to get more information about the excitation properties. 
 

All three compounds, Na$_3$Ni$_2$BiO$_6$, Li$_3$Ni$_2$SbO$_6$ and Na$_3$Ni$_2$SbO$_6$,
develop antiferromagnetic orders at low temperatures~\cite{Seibel,2019JMMM481100W}. 
Na$_3$Ni$_2$BiO$_6$
actually shows a ferromagnetic Curie-Weiss temperature, indicating competing
ferromagnetic and antiferromagnetic interaction. It is possible that the 
first neighbor $J_1$ interaction here is ferromagnetic and the second neighbor 
$J_2$ interaction is antiferromagnetic. This is captured by the model by 
flipping the spins from one sublattice. For Li$_3$Ni$_2$SbO$_6$ and Na$_3$Ni$_2$SbO$_6$,
high field and high frequency electron spin resonance measurements do provide 
decisive roles of magnetic anisotropy with polycrystal samples.

The compound Ni$_2$Mo$_3$O$_8$ cannot be regarded as a genuine spin-1 honeycomb magnet
as the two Ni sublattices experience different crystal field environments~\cite{PhysRevMaterials.3.014410,2019arXiv190602215L},
i.e. an octahedral Ni$^{2+}$ and a tetrahedral Ni$^{2+}$. Although
the spin part for both sublattices is spin-1, the tetrahedral Ni$^{2+}$
would have an active orbital degree of freedom with a partially filled 
$t_{2g}$ level, and as a result, the atomic spin-orbit coupling could then 
play an important role for the tetrahedral Ni$^{2+}$ ions and entangle
the spin and orbitals. Thus, this system is not captured by the model in our work,
and a careful modeling requires the involvement of the orbitals on one sublattice. 


Finally, we remark on the possibility of multiple-$q$ states that may emerge 
in the system with magnetic fields and/or at finite temperatures. This kind 
of states could appear in systems with strong frustration where several equivalent 
spiral $q$ wavevectors are available~\cite{PhysRevLett.108.017206}. 
In our model, multiple-$q$ states
could appear near the phase boundary between the quantum paramagnet and 
the spiral ordered side where the system is considering to pick up which $q$
wavevector or several $q$ wavevectors to generate the magnetic order. 
Likewise, as one cools the system from high temperature paramagnetic
phase on the spiral ordered side, the system would experience a similar 
frustration in choose $q$ wavevectors to generate magnetism. This effect
could persist even in the presence of external magnetic fields. 
The skyrmion lattice may be stabilized with multiple-$q$ states and magnetic 
fields. These possibilities require an optimization or variational study of 
the total energy or free energy with respect to the candidate states and are 
left for future works. 


To summarize, we have proposed a generic spin model that incorporates
the first and second neighbor exchange interactions and an
easy-plane anisotropy in this paper. We established the ground state phase diagram
as plotted in Fig.~\ref{fig1}. When the easy-plane anisotropy is much larger
compared to exchange interactions, the system lies in a quantum paramagnet
phase without any magnetic order, and the magnetic excitation develops 
a contour degenerate due to the frustration in the exchange interactions. 
When the exchange interactions are dominant, a N\'eel or spiral magnetic 
order is established at zero temperature from the quantum order by disorder. 
The disordered quantum paramagnet and the ordered N\'eel 
or spiral magnetism are separated by a frustrated quantum criticality. 
Thus, the role of frustration has been greatly extended from 
the ordered side to the disordered one and the associated magnetic 
excitations and quantum transition.

\section*{Acknowledgments}

This work is supported by research 
funds from the Ministry of Science and Technology of China with 
grant No.2018YFGH000095, No.2016YFA0301001 and No.2016YFA0300500,
and from the Research Grants Council of Hong Kong with General 
Research Fund Grant No.17303819. ZQW is supported by the U.S. 
Department of Energy, Basic Energy Sciences Grant No. DE-FG02-99ER45747.

\bibliography{Ref}

\end{document}